\documentclass[10pt,letterpaper,twocolumn]{revtex4}  

\newcommand{\ud}{\mathrm{d}}
\newcommand{\mr}{\mathbf{r}}

\newcommand{\mG}{\mathbf{G}}
\newcommand{\mk}{\mathbf{k}}
\newcommand{\mE}{\mathbf{E}}

\newcommand{\me}{\mathbf{e}}

\usepackage{graphics}
\usepackage{amsmath}

\begin{document}


\title{Light propagation in finite-sized photonic crystals:\\Multiple scattering using an electric field integral equation}


\author{Philip Tr\o st Kristensen,$^{1,*}$  Peter Lodahl$^1$ and Jesper M\o rk$^1$}
\address{$^1$DTU Fotonik, Department of Photonics Engineering, \\ Technical University of Denmark, \\\O rsteds Plads Building 343, DK-2800 Kgs. Lyngby, Denmark}
\address{$^*$Corresponding author: ptkr@fotonik.dtu.dk}

\begin{abstract}
We present an accurate, stable and efficient solution to the Lippmann-Schwinger equation for electromagnetic scattering in two dimensions. The method is well suited for multiple scattering problems and may be applied to problems with scatterers of arbitrary shape or non-homogenous background materials. We illustrate the method by calculating light emission from a line source in a finite sized photonic crystal waveguide.
\end{abstract}



\maketitle

\section{Introduction}
Design and development of future optical components rely heavily on adequate descriptions of light propagation in micro and nano structured environments. Although the governing differential equations have been known for over a century, analytical solutions are available only for a limited number of highly symmetrical problems. Therefore, in most cases of practical interest, numerical methods are employed. Ideally, the numerical method should be accurate enough to capture all relevant physics and allow for understanding the different scattering channels; yet it should be fast enough to ensure acceptable runtimes for practically relevant structures. 

The invention of photonic crystals \cite{Bykov_SovJQuantum_4_861_1975, Yablonowitch_PRL_58_2059_1987, John_PRL_58_2486_1987} has offered hitherto unprecedented control of light propagation. Consequently, these novel materials are of great importance in the design of optical components for future information technology as well as fundamental solid state quantum optics experiments. 
In the context of light propagation in micro and nano-structured media, such as photonic crystals, the full wave nature of the electromagnetic field needs to be taken into account. A large number of different methods are being explored in the investigation of light propagation in these materials, including plane wave expansion for band structure calculations \cite{Leung_PRL_65_2646_1990}, the Generalized Multipole Technique \cite{Hafner} and the Finite Difference Time Domain (FDTD) method \cite{Tavlove}. In particular, we note that for calculations involving large numbers of scatterers in an otherwise homogeneous background, Rayleigh-multipole methods have been used for calculations on microstructured fibres \cite{White_JOSAB_19_2002, White_JOSAB_19_2002_errata, Boris_JOSAB_19_2002, Campbell_JOSAB21_1919_2004} as well as photonic crystals composed of cylinders \cite{Asatryan_Waves_in_Random_Media_13_9_2003, Fussel_PRE_70_066608_2004} or spheres \cite{Wang_PRB_47, Yannopapas_PRB_75_115124_2007}. The use of multipole expansions for the fields ensures a significant reduction in the number of basis functions, thus enabling calculations on complex structures of practical interest.

In addition to the above methods, various volume and surface integral methods exist for the electric and/or magnetic fields. These methods typically rely on discretization to express the integrals as linear systems of equations - a procedure known as the method of moments. In the discretization process the solution is expanded in terms of linearly independent basis functions which vary in complexity, depending on the specific method employed \cite{PetersonRayMittra}. A particularly simple and convenient choice of basis function is the pulse basis function resulting in what is usually termed the coupled (or discrete) dipole approximation (CDA) \cite{Purcell_AstrophysJ_186_705_1973, Draine_JOSAA_4_1491_1994}. The CDA allows for relatively easy implementation as well as the physically attractive property that the resulting field can be directly interpreted as the sum of the field from all the individual cells in the scattering structure oscillating as dipoles in response to the incoming field. Depending on the desired accuracy and the nature of the scattering problem at hand, however, the required number of basis functions may become too large for practical calculations on e.g. photonic crystals. In addition, this type of discretization may lead to stability problems which, for example, limit the efficiency of volume integral equations for the electric field in the case of high refractive index contrasts \cite{PetersonRayMittra}.

For use in the modelling of nanophotonic structures and in particular in the context of light-matter interaction in photonic crystals, we focus in this paper on an integral type scattering formulation of light propagation based on the electric field Green's tensor. The Green's tensor, $\mG(\mathbf{r},\mathbf{r}')$, is the electromagnetic propagator which, for a given system, holds all information necessary to solve the inhomogeneous vector Helmholtz equation and may be interpreted as the electric field at point $\mathbf{r}$ due to an oscillating point dipole at the point $\mathbf{r}'$. Of special importance in nanophotonics modelling is the imaginary part of the Green's tensor at $\mr=\mr'$ which is proportional to the local optical density-of-states (LDOS) \cite{Sprik_EuroPhys_35_265_1996}. The LDOS and/or the Green's tensor for photonic crystals have previously been investigated theoretically using plane wave expansions \cite{John_PRE_58_3896_1998, Wang_PRB_67_155114_2003} and FDTD \cite{Koenderink_JOSAB_23_1196_2006, MangaRao_PRB_75_205437_2007} as well as through an expansion in eigenstates \cite{CowanYoung_PRE_68_046606_2003,Hughes_OL_30_1393_2005}. The application of the CDA to the calculation of the Green's tensor in a photonic crystal slab was reported in Ref. \cite{Zhao_PRE_72_026614_2005}. Using the Green's tensor for the background medium, one may recast the wave equation as a scattering problem, in which case the solution is given in terms of the so-called Lippmann-Schwinger integral equation \cite{Levine_Schwinger_PR_74_958_1948}. Since the Green's tensor is known for a number of simple material configurations, these may be used as the background material in order to limit the extend of the numerical calculations. In particular, expressions for the Green's tensor for stratified media are available \cite{Paulus_PRE_62_2000,Paulus_PRE_63_2001}.


In this article we describe a multiple scattering solution to the Lippmann-Schwinger equation. The method is developed in two dimensions, but we note that a similar approach is possible for three dimensional problems as well. The method will find applications in modelling of nanophotonic structures and devices, such as waveguides, junctions and filters as well as switches and single photon sources based on photonic crystals. In light of the above discussion the method may be viewed as a hybrid between integral type method of moments calculations and multiple scattering multipole methods. In our approach, the Lippmann-Schwinger equation is first expanded in cylinder functions (so-called normal modes) and solved within the scattering elements. Solutions at points outside the scattering elements are subsequently calculated directly using the Lippmann-Schwinger equation. Because of the integral formulation the method may benefit from known results for the Green's tensor in the background material while the normal mode expansion reduces the number of basis functions needed, thus enabling calculations on material systems of practical relevance. In addition, we make use of a number of theorems which are regularly employed in multipole methods in order to simplify the evaluation of the scattering matrix elements.

The article is organized as follows: Section \ref{Sec:Formulation} introduces the method and provides a number of results that may be of value in the practical implementation. The calculations are illustrated in section \ref{Sec:ExampleCalculations} through the solution of two small scattering problems and in section \ref{Sec:AccuracyOfTheMethod} we use these as examples to discuss a practical method of evaluating the accuracy of a given calculation. In section \ref{Sec:ApplicationExample} we provide an example application of the method in the form of a two-dimensional photonic crystal waveguide at the edge of a dielectric block. Section \ref{Sec:Conclusion} gives the conclusions.


\section{Formulation of scattering problem and solution method}
\label{Sec:Formulation}
We consider scattering of monochromatic light, $\mE(\mr,t)=\mE(\mr)\exp(-i\omega t)$ in two dimensions, corresponding to scattering problems in which the geometry is invariant along the $z$-direction and the light travels in the $xy$-plane. We limit the discussion to non-magnetic materials and consider general material systems, consisting of a finite number of piecewise constant dielectric perturbations to a piecewise homogeneous background. This is the case, for example, for most photonic crystals. The electric field $\mE(\mr)$ solves the vector wave equation
\begin{equation}
\nabla\times\nabla\times\mE(\mr) - k_0^2\epsilon(\mr)\mE(\mr) = 0,
\label{Eq:GeneralWaveEquation}
\end{equation}
where $\epsilon(\mr)$ is the position dependent permittivity and $k_0=|\mk_0|=\omega/c$ is the wave number in vacuum. For two-dimensional problems, the vector equation decouples into two independent equations for the Transverse Electric (TE) and the Transverse Magnetic (TM) polarizations. In the case of TE polarization, the electric field is oriented in the $xy$-plane, whereas, for TM polarization the electric field is oriented along the $z$-axis and the scattering calculation is essentially a scalar problem.
%
%
In order to reformulate Eq. (\ref{Eq:GeneralWaveEquation}) as a scattering problem, we consider the change in permittivity, $\Delta\varepsilon(\mr)=\epsilon(\mr)-\epsilon_B(\mr)$, caused by the introduction of scattering sites into the background medium described by $\epsilon_B(\mr)$. The solution to Eq. (\ref{Eq:GeneralWaveEquation}) with $\epsilon(\mr)=\epsilon_B(\mr)$ is denoted by $\mE^B(\mr)$ and represents the incoming field. The full solution to Eq. (\ref{Eq:GeneralWaveEquation}) is the sum of the incoming field and the scattered field. It is given by the Lippmann-Schwinger equation
\begin{equation}
\mE(\mr)=\mE^B(\mr)+\int_D\mG^B(\mr,\mr')\,k_0^2\Delta\varepsilon(\mr')\,\mE(\mr')\ud \mr',
\label{Eq:LippmannSchwinger}
\end{equation}
in which $\mG^B(\mr,\mr')$ is the Green's tensor for the background medium. For a homogeneous background, $\epsilon_B(\mr) = \epsilon_B=n_B^2$, the two dimensional Green's tensor is given as  \cite{Martin_PRE58_3909_1998}:
\begin{equation}
G ^B_{2D}(\mr,\mr')=\left(1+\frac{\nabla\nabla}{k_B^2}\right)\frac{i}{4}H_0(k_B|\mr-\mr'|)
\label{Eq:G2DdifferentialForm}
\end{equation}
in which $k_B=n_Bk_0$ is the wave number in the background medium and $H_0(r)$ is the Hankel function of the first kind of order $0$.

\subsection{Expansion in normal modes}
In two and three dimensions the real part of the Green's tensor diverges in the limit $\mr=\mr'$, which means, that for integrals in which $\mr$ is inside the scattering volume (such as in this work) an alternative formulation of the Lippmann-Schwinger equation must be employed in which the singularity is isolated in an infinitesimal principle volume and treated analytically \cite{Yaghjian_proc_IEEE_68_248_1980}. Therefore, we follow Ref. \cite{Martin_PRE58_3909_1998} and rewrite the equation as
\begin{align}
\mE(\mr)=\mE^B(\mr)&+\lim_{\delta V\rightarrow 0}\int_{D-\delta V}\mG^B(\mr,\mr')\,k_0^2\Delta\varepsilon(\mr')\mE(\mr')\ud \mr' \nonumber \\
&- \mathbf{L}\frac{\Delta\epsilon(\mr)}{\epsilon_B} \mE(\mr),
\end{align}
in which $L_{xx}=L_{yy}=1/2$ and $L_{\alpha\beta}=0$ otherwise, corresponding to a circular exclusion volume $\delta V$ centered on $\mr'=\mr$.

We will solve for the total field, $\mE(\mr)$, inside the scattering material only, as the solution everywhere else can be subsequently obtained by use of the Lippmann-Schwinger equation. The incoming field, $\mE^B(\mr)$, is a solution to the wave equation with no scatterers and thus, in general, can be expanded on the solutions in the bulk background material. Similarly, the total field at positions inside each scatterer may be expanded on the solutions in a homogeneous material with the same permittivity. Based on these considerations we construct a basis consisting of normal modes, each with support on only one of the scattering sites:
\begin{align}
\psi_n &= K_nJ_{q}(k_dr_d)e^{iq\varphi_d}\,S_d(\mr) \nonumber \\
\psi^B_n &= K^B_nJ_{q}(k_Br_d)e^{iq\varphi_d}\,S_d(\mr) \nonumber,
\end{align}
where we have defined a combined index $n=(q,d,\alpha)$, in which $q$ denotes the order of the normal mode and $d$ denotes the particular subdomain $D_d$ of the scattering material we consider, e.g. which cylinder. For convenience we have included the field component $\alpha\in\{x,y,z\}$ in the index $n$ as well. $J_q(r)$ denotes the Bessel function of the first kind of order $q$ and $k_d=n_dk_0$, where $n_d$ denotes the refractive index in the subdomain $D_d$. ($r_d$,$\varphi_d$) denote polar coordinates with respect to the local coordinate system in the subdomain $D_d$ and $S_d=1$ for $\mr \in D_d$ and $S_d=0$ otherwise. $K_n$ and $K ^B_n$ are normalization constants. To ease the notation we will write $n$ in place of any of the indices $q,d,\alpha$ and we will supress the index on the coordinates $(r,\varphi)$ as this will not lead to ambiguities.

We define an inner product as
\begin{equation}
\langle\psi_m|\psi_n\rangle=\int \psi_m^*(\mr)\,\psi_n(\mr)\,\ud \mr,
\label{Eq:Delta_mn}
\end{equation}
which is in general non-zero for $m\neq n$, and we normalize the basis functions so that $\langle\psi_n|\psi_n\rangle=\langle\psi ^B_n|\psi^B_n\rangle=1$. By expanding the electric fields as
\begin{align}
\mE(\mr)&=\sum_{n}e_n\psi_n(\mr)\me_n \nonumber\\
\mE^B(\mr)&=\sum_{n}e^B_n\psi^B_n(\mr)\me_n, \nonumber
\end{align}
where $\me_n$ is a unit vector in the direction $\alpha$, and projecting onto the basis formed by $\psi_m\me_m$ and $\psi ^B_m\me_m$, the Lippmann-Schwinger equation may be rewritten in matrix form as
\begin{align}
\Big(1+L_{mn}\frac{\Delta\epsilon}{\epsilon_B}\Big)&\sum_{n}\langle\psi_m|\psi_n\rangle e_n = \sum_{n}\langle\psi_m|\psi^B_n\rangle e^B_n \nonumber \\
&+ k^2\Delta\varepsilon\sum_{n} G^{\alpha\beta}_{mn}e_n
\label{Eq:GeneralEqSystem}
\end{align}
in which
\begin{align}
& G^{\alpha\beta}_{mn} =\int_D\psi_m^*(\mr)\lim_{\delta V\rightarrow 0}\int_{D-\delta V}G^B_{\alpha\beta}(\mr,\mr')\psi_n(\mr')\ud\mr'\ud\mr,
\label{Eq:MatrixElementsGeneral}
\end{align}
and we have written explicitly the directions $\alpha,\beta$ corresponding to the indices $m,n$ for clarity.

Eqs. (\ref{Eq:G2DdifferentialForm}) - (\ref{Eq:MatrixElementsGeneral}) hold the complete formulation of the method. The form of the central matrix equation (\ref{Eq:GeneralEqSystem}) is only slightly different from that of the CDA in that the left hand side is generally a matrix (although this matrix is very sparse, since basis functions belonging to different scattering domains are orthogonal by construction). The underlying strategy and the practical implementation of the two methods, however, are very different. In the CDA, the projection onto the basis functions is straight forward, resulting typically in a very large matrix equation system that is subsequently solved by iterative methods. In the present method it is the projections onto the basis functions that are potentially time consuming, but eventually leads to a relatively small system of equations. A number of mathematical results exist, though, that can dramatically speed up the calculations of the matrix elements and the implementation in general. In sections \ref{Sec:SelfTerms}-\ref{Sec:ExteriorSolution} we discuss these results.

The matrix elements need to be evaluated for basis functions within the same domain (self-terms) as well as different domains (scattering terms). The case of a homogeneous background is of special interest as one will often be able to separate the background Green's tensor into terms corresponding to a homogeneous background and a number of additional scattering terms. Therefore, we focus in this section on the evaluation of the matrix elements for a homogeneous background Green's tensor, Eq. (\ref{Eq:G2DdifferentialForm}), and return to the additional terms due to scattering in an inhomogeneous background in section \ref{Sec:ApplicationExample}.

\subsection{Self-terms}
\label{Sec:SelfTerms}
The evaluation of the self-terms is complicated by the divergence in the Green's tensor for $\mr'=\mr$ and the fact that the integrand couples $\mr'$ and $\mr$, effectively resulting in a four-dimensional integral. In the following we adopt a particular method for the evaluation, in which the four-dimensional integral, Eq. (\ref{Eq:MatrixElementsGeneral}), is rewritten in terms of a number of one and two dimensional integrals. Although this method is suitable for evaluation of general matrix elements, we note that other methods may be more efficient in case of a particular geometry and/or polarization.
\begin{figure}
\centering
\includegraphics{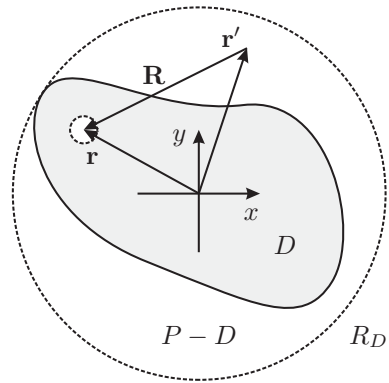}
\caption{\label{Fig:IntegrationSelfTerm} Sketch of the local coordinates used for the calculation of the self-term in scattering domain $D$. }
\end{figure}

Figure \ref{Fig:IntegrationSelfTerm} shows a sketch of the local coordinates used for the evaluation of the integral. In order to efficiently treat the divergence, for each $\mr$ we first integrate $\mr'$ over the entire plane $P$ less the principal volume centered on $\mr$. Subsequently, we subtract the integral for $\mr'\notin D$, for which the limit $\delta V\rightarrow 0$ is trivial, since $\mr'\neq\mr$. The matrix element is thus rewritten as $G^{\alpha\beta}_{mn} = \mathcal{A}^{\alpha\beta}_{mn} - \mathcal{B}^{\alpha\beta}_{mn}$, in which
\[
\mathcal{A}^{\alpha\beta}_{mn} = \int_D\psi^*_m(\mr) \lim_{\delta V\rightarrow 0}\int_{P-\delta V} G^B_{\alpha\beta}(\mr,\mr')\psi_n(\mr')\ud \mr' \ud\mr
\]
and
\[
\mathcal{B}^{\alpha\beta}_{mn} = \int_D\psi^*_m(\mr) \int_{P-D} G^B_{\alpha\beta}(\mr,\mr')\psi_n(\mr')\ud \mr' \ud\mr.
\]

Using the Graf addition theorem (cf. appendix \ref{Sec:AdditionTheorems}), we can simplify the evaluation of $\mathcal{A}^{\alpha\beta}$ by expanding the function $\psi_n(\mr')$ around the point $\mr$, so that

\begin{align}
\mathcal{A}^{\alpha\beta} &= \sum_\mu\int_D J_{m}(k_Rr)J_{n+\mu}(k_Rr)e^{i(n+\mu-m)\varphi} \nonumber \\
&\quad\times\int_{P-\delta V}G^B_{\alpha\beta}(\mr,\mathbf{R})J_\mu(k_RR)(-1)^{\mu}e ^{-i\mu\theta} \ud \mathbf{R} \ud\mr \nonumber \\
&= \sum_\mu\int_D J_{m}(k_Rr)J_{n+\mu}(k_Rr)e^{i(n+\mu-m)\varphi} \ud\mr \times I^{\alpha\beta}_{\mu}, \nonumber
\end{align}
where $k_R=n_dk_0$ and we have exploited the fact that the integration over $\mathbf{R}$ is over the entire plane (less the principal volume), and thus does not depend on $\mr$. The evaluation of this integral over $\mathbf{R}$, denoted by $I^{\alpha\beta}_\mu$ above, can be reduced to a number of one dimensional integrals as shown in appendix \ref{Sec:MatrixElementCalculations}. We note that the simple angular dependence of the integrands in many cases allows for a reduction of the remaining integral over $\mr$ to a sum of one dimensional integrals.

Evaluation of $\mathcal{B}^{\alpha\beta}_{mn}$ may also be substantially simplified using the Graf addition theorem to expand the Hankel function in terms of Bessel and Hankel functions defined in the local coordinate system. The expansion differs depending on the sign of $r-r'$; for $r'>r$ we write the integrand as
\begin{align}
{\mathit{b}}^{\alpha\beta}_{mn}(\mr,\mr') &= \psi^*_m(\mr)G^B_{\alpha\beta}(\mr,\mr')\psi_n(\mr')\nonumber \\
&= \sum_\mu J_{m}(k_Rr)e^{-im\varphi} \mathcal{L}^{\alpha\beta}\left\{\frac{i}{4}J_\mu(k_B r)e^{-i\mu\varphi}\right\} \nonumber \\
&\quad\times H_\mu(k_Br')J_{n}(kRr'e^{i(n+\mu)\varphi'}),
\label{Eq:bRpGtR}
\end{align}
whereas, for $r'<r$ we write
\begin{align}
{\mathit{b}}^{\alpha\beta}_{mn}(\mr,\mr') &=  \sum_\mu J_{m}(k_Rr)e^{-im\varphi} \mathcal{L}^{\alpha\beta}\left\{\frac{i}{4}H_\mu(k_B r)e^{i\mu\varphi}\right\} \nonumber \\
&\quad\times J_\mu(k_Br')J_{n}(kRr'e^{i(n-\mu)\varphi'}),
\label{Eq:bRGtRp}
\end{align}
in which $\mathcal{L}^{\alpha\beta}$ is the $\alpha,\beta$ component of the linear operator in Eq. (\ref{Eq:G2DdifferentialForm}). Derivatives for the general cylinder functions $\partial_\alpha\partial_\beta\{Z_\mu(k_B r)e^{\pm\,i \mu\varphi}\}$ are provided in appendix \ref{Sec:GeneralDerivatives}. For circular scatterers we always have  $r'>r$ and the expression for $\mathcal{B}^{\alpha\beta}_{mn}$ factors into a number of one dimensional integrals. Similarly, the evaluation of $\mathcal{B}^{\alpha\beta}_{mn}$ for non-circular scatterers may be conveniently split depending on whether $\mr'$ is outside or inside the circumscribing circle (denoted by $R_D$ in Fig. \ref{Fig:IntegrationSelfTerm}). In the former case, the expression factors into separate integrals for $\mr$ and $\mr'$, whereas in the latter case, the two integrations are coupled. Again, the simple angular dependence of the integrands in many cases allows for a reduction of these integrals to a sum of two dimensional integrals.

\subsection{Scattering terms}
\label{Sec:ScatteringTerms}
\begin{figure}[htb]
\centering
\includegraphics{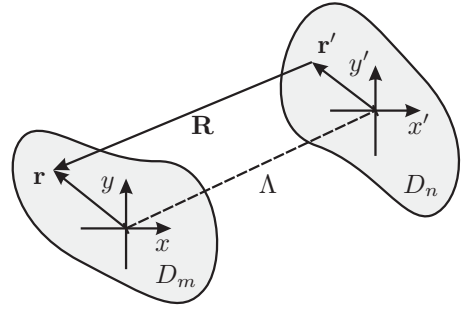}
\caption{\label{Fig:IntegrationScatteringTerm}Sketch of local coordinates for $\mr$ and $\mr'$ in two independent scatterers. Using the Graf addition theorem the calculation of the scattering matrix elements may be simplified considerably and expressed in terms of the integral over the individual scatterers and the distance $\Lambda$ between the centers of the two local coordinate systems.}
\end{figure}

For the calculation of scattering terms the integration domains for $\mr$ and $\mr'$ are completely separated in space and so the Green's tensor is well behaved at all points of interest. In this case we employ the Graf addition theorem twice to express the Hankel function in terms of the distance between the centers of the two local coordinate systems, as illustrated in Fig. \ref{Fig:IntegrationScatteringTerm}
\begin{align}
G^{\alpha\beta}_{mn} = &\frac{i}{4}\sum_{\mu,\lambda}H_{\mu+\lambda}(k_B \Lambda)\int_{D_m} J_{m}(k_mr)e^{-i\,m\,\varphi}\nonumber \\
&\times\mathcal{L}^{\alpha\beta}\left\{J_{\lambda}(k_B r)e^{-i \lambda\varphi}\right\}\ud\mr\nonumber \\
&\times\int_{D_n} J_\mu(k_Br')J_{n}(k_nr')e^{i\,(n-\mu)\,\varphi'}\ud\mr'.
\label{Eq:GzzScatteringtermGrafForm}
\end{align}
Eq. (\ref{Eq:GzzScatteringtermGrafForm}) shows that the scattering matrix calculation factors into terms that depend only on the geometries of the individual scatterers and the distance between them. Since the Hankel function as well as the Bessel functions are well behaved at all points of interest, the integrals may be directly evaluated for arbitrary geometries, e.g. using numerical quadrature. Note that the procedure outlined above is compromised when $\Lambda<R_m+R_n$, where $R_m$ and $R_n$ are the radii of the circumscribing circles of domains $D_m$ and $D_n$, respectively. This could happen in the case of close non-circular scatterers. In this case the Graf addition theorem is not valid and one will need to employ a strategy similar to Eqs. (\ref{Eq:bRpGtR}) and (\ref{Eq:bRGtRp}).

\subsection{Background electric field}
\label{Sec:BGelectricField}
The incident background electric field, $\mE^B(\mr)$ is a solution to the wave equation without the scatterers. In the case of a bulk background, the solutions are plane waves, and the expansion in terms of cylinder functions is readily obtained using the Jacobi-Anger identity, as discussed in appendix \ref{Sec:AdditionTheorems}. Instead of using plane waves as the background electric field we may use the columns of the 2D Green's tensor. These are related to the electric field at $\mr$ due to a line source at $\mr'$ \cite{Martin_PRE58_3909_1998}. By comparing with Dyson's equation
\begin{align}
\mG(\mr,\mr') = &\mG^B(\mr,\mr')\nonumber \\
&+\int_V\mG^B(\mr,\mr'')\,k_0^2\Delta\varepsilon(\mr'')\,\mG(\mr'',\mr')\ud \mr'', \nonumber
\label{Eq:DysonEquation}
\end{align}
we see that the solution to the Lippmann-Schwinger equation in this case exactly produces the corresponding columns of the full 2D Green's tensor for the scattering problem.

\subsection{Exterior solution}
\label{Sec:ExteriorSolution}
The matrix equation (\ref{Eq:GeneralEqSystem}) is solved using standard linear algebra routines to yield the solution at any point inside the scattering domains. The solution at any point outside the scatterers can be subsequently obtained directly from the Lippmann-Schwinger equation which is now an explicit equation:
\begin{equation}
\mE(\mr)=\mE^B(\mr)+\int\mG^B(\mr,\mr')\,k_0^2\Delta\varepsilon(\mr')\sum_{n}e_n\psi_n(\mr')\me_n\ud \mr'.
\label{Eq:LippmannSchwingerMultipoleForm}
\end{equation}

The sum in Eq. (\ref{Eq:LippmannSchwingerMultipoleForm}) runs over all basis functions in all scattering domains. Again, the calculation may be simplified considerably by the use of the Graf addition theorem to rewrite the equation in terms of the distances from the centers of the local coordinate systems. Considering for simplicity the case of just a single scattering domain $D$ we rewrite the equation as
\begin{align}
\mE(\mr)=&\mE^B(\mr)+\frac{i}{4}\Delta\varepsilon_dk_0^2\sum_{\mu,n}\, \mathcal{L}\left\{H_\mu(k_B L)e^{i\mu\theta}\right\}(-1)^{\mu}\me_n \nonumber \\
&\times\int_D J_\mu(k_B r')e_nK_nJ_{n}(k_Rr')e^{i\,(n-\mu)\,\varphi'}\ud\mr',
\label{Eq:LippmannSchwingerGrafForm}
\end{align}
where now $(L,\theta)$ are polar coordinates of $\mathcal{O}'$ with respect to $\mr$, cf. appendix \ref{Sec:AdditionTheorems}.

\subsection{Example calculations}
\label{Sec:ExampleCalculations}

\begin{figure}[b]
\centering
\includegraphics{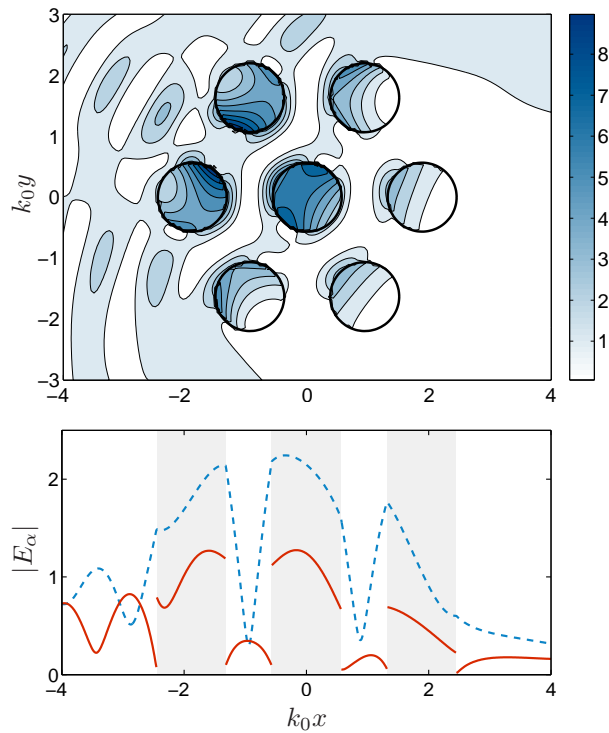}
\caption{\label{Fig:airHolesInGaAsHex2DPC}Example calculation. A TE plane wave of unit amplitude, $\mE^B(\mr)=\me\exp(i\,n_B\mk_0\cdot\mr)$, is incident from the top left on a crystallite consisting of seven air holes ($n_d=1$) in a high-index dielectric background ($n_B=3.5$). Parameters are $\mk_0=(\sqrt{3}/2,-1/2)$ and $R=0.3a$ where $R$ is the radius of the cylindrical holes and $a=0.3\lambda_0$ is the distance in between. Top: Absolute square, $|\mE(\mr)|^2$, of the resulting field as a function of position in the xy plane. Bottom: Absolute value of the components $E_x(x)$ (red solid line) and $E_y(x)$ (blue dashed line) along the line $y=0$.}
\end{figure}

To illustrate the method we consider now an example scattering problem in which a TE plane wave is incident from the top left on a small crystallite of air cylinders in a high-index dielectric. Fig. \ref{Fig:airHolesInGaAsHex2DPC} shows the absolute square of the total field as a function of the position in the $xy$-plane. Also we show the magnitude of the $E_x$ and the $E_y$ components of the field along the line $y=0$ through the centres of three of the cylinders. Clearly, the x component shows a number of discontinuous jumps, whereas the y component is continuous in accordance with the boundary conditions. We note that the multiple scattering from the air cylinders act to partly block the light, resulting in the formation of a standing wave in the upper left part of the figure. Typically we use the same number of basis functions in each scattering domain and for each polarization, so that $|q|\leq M_{max}$. This calculation was performed using $M_{max}=10$, resulting in a matrix equation system of 294 unknowns.

\begin{figure}[htb]
\centering
\includegraphics{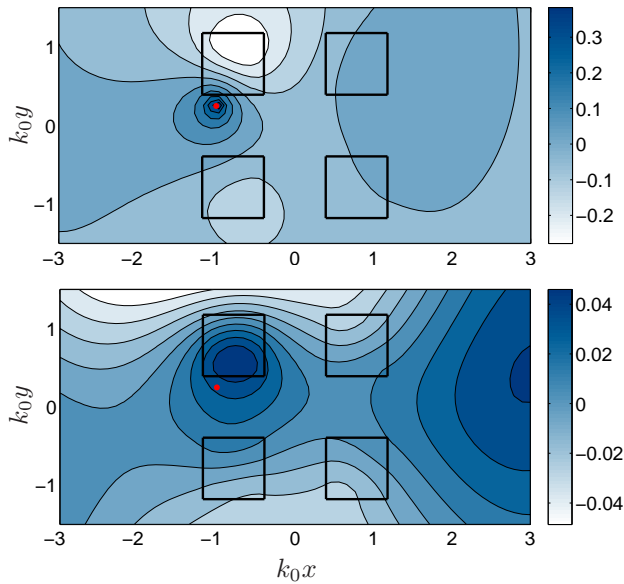}
\caption{\label{Fig:squareDielectricRods2DPC}Real (top) and imaginary (bottom) part of the total TM Green's tensor $G_{zz}(\mr,\mr')$ as a function of $\mr$ with $k_0\mr'=(-1,1/4)$ (as indicated by the red dot) in a structure consisting of four dielectric rods ($n_d=3.5$) of square cross section in air. Parameters are $a=2L$ where $a$ is the distance between the rods and $L=\lambda_0/4$ is the side length.}
\end{figure}

As noted in section \ref{Sec:BGelectricField} we may use the present method to calculate the Green's tensor for the scattering structure. In Fig. \ref{Fig:squareDielectricRods2DPC} we consider a geometry consisting of four square dielectric rods in air and we show the real and imaginary part of the TM Green's tensor $G_{zz}(\mr,\mr')$ as a function of $\mr$ for constant $\mr'$ indicated in the figure. The real part diverges in the limit $\mr\rightarrow\mr'$, whereas the imaginary part is continuous at all points. In the limit $\mr=\mr'$ it is proportional to the LDOS, as noted in the introduction. The calculaton was performed using $M_{max}=10$, resulting in only 44 unknowns.

\section{Accuracy of the method}
\label{Sec:AccuracyOfTheMethod}
\begin{figure}[b]
\centering
\includegraphics{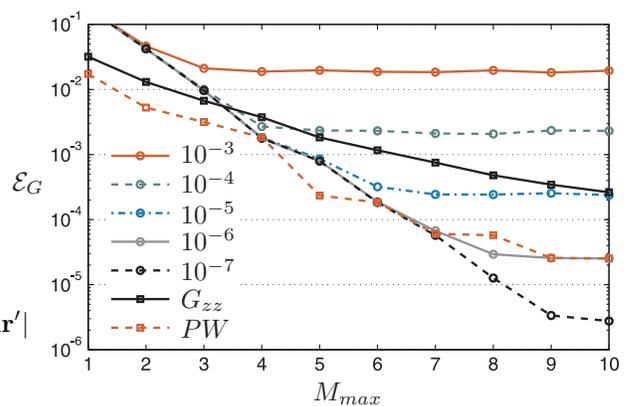}
\caption{\label{Fig:airHolesInGaAsHex2DPC_ErrorAnalysis}Global error as a function of the number of basis functions used in the expansion of the electric fields (controlled by $M_{max}$). Circular markers correspond to the problem in Fig. \ref{Fig:airHolesInGaAsHex2DPC} with different curves corresponding to different fixed errors on the relevant matrix elements as indicated. Square markers corresond to the problem in Fig. \ref{Fig:squareDielectricRods2DPC} calculated for the Green's tensor ($G_{zz}$) and plane waves ($PW$) as the background field.}
\end{figure}

The numerical error stems primarily from evaluation of the matrix elements and the truncation of the basis set. After solving the linear equation system, Eq. (\ref{Eq:GeneralEqSystem}), the accuracy of a given solution may be estimated by substitution back into the Lippmann-Schwinger equation. To this end we define the local error as
\begin{align}
\mathcal{E}_L(\mr) =| \mE^B(\mr)-\mE_{num}(\mr)+\int\mG^B(\mr,\mr')\,k_0^2\Delta\varepsilon(\mr')\,\mE_{num}(\mr')\ud \mr'|
\label{Eq:RelError}
\end{align}
and we note that since $\mE^B(\mr)$ and $\mG^B(\mr,\mr')$ are known analytically, we can use this as a measure of the accuracy of a given solution even if we do not know the analytical solution. Based on the local error, we define the global relative error as
\[
\mathcal{E}_G=\frac{\int\mathcal{E}_L(\mr)\ud\mr}{\int|\mE^B(\mr)|\ud\mr},
\]
where the integrals are taken over the area of the scattering sites only. Fig. \ref{Fig:airHolesInGaAsHex2DPC_ErrorAnalysis} shows the global error as a function of the number of basis functions used in the expansions and dependent on the error in the matrix elements for the solutions depicted in Figs. \ref{Fig:airHolesInGaAsHex2DPC} and \ref{Fig:squareDielectricRods2DPC}. The error analysis was performed by first calculating the matrix elements to high precision, using an absolute error tolerance on the numerical integrals of $10^{-6}$. Subsequently, for each value of $M_{max}$ the corresponding linear equation system, Eq. (\ref{Eq:GeneralEqSystem}) was constructed and a random complex number of fixed modulus, $\delta G_{mn}$ was added to each element in the matrix of modulus larger than $\delta G_{mn}$ before solving the equation system.

The analysis shows an exponential like decrease in the global error as a function of the number of basis functions, underscoring the massive reduction in basis functions due to the expansion in normal modes when compared to conventional discretization methods. This is the case for the cylindrical holes in Fig. \ref{Fig:airHolesInGaAsHex2DPC} as well as for the  square rods in Fig. \ref{Fig:squareDielectricRods2DPC}. The convergence is faster in the case of the cylindrical holes, which is partly because the basis functions have the same symmetry as the scatterers and partly because the plane wave field is easier to approximate than the (divergent) Hankel function. Clearly, the artificial error on the matrix elements acts to limit the minimum achievable global error, and the analysis thus confirms that the global error is controlled by the number of basis functions as well as the accuracy of the numerical quadrature. We note, that the measure, Eq. (\ref{Eq:RelError}), may be viewed as a test of self-consistency of the method which is of principal importance for any solution to Eq. (\ref{Eq:LippmannSchwinger}). From Fig. \ref{Fig:airHolesInGaAsHex2DPC_ErrorAnalysis} we can see that the measure is also of practical importance, since, for a given tolerance on the numerical integrals, it can be used to estimate the number of basis functions needed to reach the minimum global relative error.


\section{Example application: Light emission in a finite sized photonic crystal waveguide}
\label{Sec:ApplicationExample}
As an example of the utility of the method we present in this section results for the investigation of light propagation near the edge of a finite sized two dimensional photonic crystal. We consider a photonic crystal made from 80 circular rods of refractive index $n_d=3.4$ in a lower-index background ($n_B=1.5$). The cylinders are placed in a square lattice, and a short waveguide is created by the omission of 4 rods along the (11) direction of the crystal. The waveguide along with the crystal is terminated by an interface to air. We focus on TM polarized light and calculate the Green's tensor of the system $G_{zz}(\mr,\mr')$. Although the integral expressions become larger, a similar procedure as the one outlined below may be used for the calculation of TE polarized light as well as for multiple interfaces. We start by extending the formalism of the previous sections to the case of a non-homogeneous background Green's tensor in section \ref{Sec:AdditionalScattering} and go on to show example calculations of light emission from a finite sized photonic crystal in section \ref{Sec:PhotonGunTM}.

\subsection{Additional scattering near interface}
\label{Sec:AdditionalScattering}
For the scattering calculations near a dielectric interface we use the Green's tensor for the dielectric half-space as the background Green's tensor in Eq. (\ref{Eq:LippmannSchwinger}). The 2D Green's tensor for general strattified media is given in Ref. \cite{Paulus_PRE_63_2001}. It is expressed in terms of an integral in $\mk$-space and below we discuss the calculation of the elements $G^{zz}_{mn}$ in the special case of a single dielectric interface. We consider TM polarized light incident on an interface at $y=0$ between two media with refractive indices $n_A$ and $n_B$, respectively. We will deal only with scatterers in the lower layer (layer B) and so, following Ref. \cite{Paulus_PRE_63_2001}, the 2D Green's tensor is given as
\begin{align}
G^B_{zz}(\mr,\mr') &= -\frac{\mathbf{\hat{y}}\mathbf{\hat{y}}}{k_2^2}\delta(\mathbf{R}) \nonumber \\
&+ \frac{i}{4\pi}\int_{-\infty}^{\infty}\frac{1}{k_{B,y}}e^{i\,k_x(x-x')}e^{i\,k_{B,y}|y-y'|}\ud k_x \nonumber \\
&+ \frac{i}{4\pi}\int_{-\infty}^{\infty}\frac{\mathcal{F}^S_{BA}}{k_{B,y}}e^{i\,k_x(x-x')}e^{-i\,k_{B,y}(y+y')}\ud k_x,
\label{Eq:2DGreensInterfaceTM}
\end{align}
where $k_{l,y}=\sqrt{k_l^2-k_x^2}$ with $k_l=n_lk_0, l\in\{A,B\}$ and
\[
\mathcal{F}^S_{BA} = \frac{k_{B,y}-k_{A,y}}{k_{B,y}+k_{A,y}} = \frac{\sqrt{k_B^2-k_x^2}-\sqrt{k_A^2-k_x^2}}{\sqrt{k_B^2-k_x^2}+\sqrt{k_A^2-k_x^2}}
\]
is the Fresnel reflection coefficient.

In Eq. (\ref{Eq:2DGreensInterfaceTM}) the first two terms correspond to the Green's tensor of the homogeneous material whereas the last term gives the reflection off the interface. This means that the evaluation of the matrix element $G^{zz}_{mn}$ naturally splits into a direct homogeneous material part and an indirect interface scattering part. The former is exactly  what was handled in sections \ref{Sec:SelfTerms} and \ref{Sec:ScatteringTerms} so we concentrate in this section only on the scattering contribution $G^S_{mn}$:
\begin{align}
G^{S}_{mn}&=\frac{i}{4\pi}\int_{-\infty}^{\infty}\frac{\mathcal{F}^S_{BA}(k_x)}{k_{B,y}(k_x)}\int_{D_m}\int_{D_n}\psi_m^*(\mr)e^{i\,k_x(x-x')}
\nonumber \\
&\times e^{-i\,k_{B,y}(k_x)(y+y')}\psi_n(\mr')\ud \mr' \ud \mr \ud k_x.
\end{align}

In order to carry out the integration we first write $(x,y)=(X,Y)+(r\cos\varphi,r\sin\varphi)$ and $(x',y')=(X',Y')+(r'\cos\varphi',r'\sin\varphi')$, where $(X,Y)$ and $(X',Y')$ denote absolute coordinates of the centers of the local coordinate systems. We then recast the expression in terms of local coordinates as
\begin{align}
G^{S}_{mn}&=\frac{i}{4\pi}\int_{-\infty}^{\infty}\frac{\mathcal{F}^S_{BA}(k_x)}{k_{B,y}(k_x)}
e^{i(k_x(X-X')-k_{B,y}(Y+Y'))} \nonumber \\
&\times \int_{D_m} J_{m}(k_mr)e^{-i\,m\,\varphi} e^{i\,k_B\,r\,\cos(\varphi-\theta)} \ud\mr\nonumber \\
&\times \int_{D_n} J_{n}(k_nr')e^{i\,n\,\varphi'} e^{i\,k_B\,r'\,\cos(\varphi'-\theta')} \ud\mr' \ud k_x,
\end{align}
where we have rewritten the inner products of the wave vectors and the position vectors in the two domains in terms of the angles between them. This angle becomes imaginary whenever $k_x^2>k_B^2$. As in the case of the homogeneous background we are able to simplify the expression further by factoring out the integrals over the domains $D_m$ and $D_n$. To this end we use the Jacobi Anger identity, cf. appendix \ref{Sec:AdditionTheorems}, to rewrite the matrix elements as
\begin{align}
G^{S}_{mn}&=\frac{i}{4\pi}\sum_{\lambda,\gamma} i^{\lambda+\gamma}\nonumber \\
&\times\int_{-\infty}^{\infty}\frac{\mathcal{F}^S_{BA}(k_x)}{k_{B,y}(k_x)}
e^{i(k_x(X-X')-k_{B,y}(Y+Y'))}e^{-i(\lambda\theta+\gamma\theta')} \ud k_x \nonumber \\
&\times \int_{D_m} J_{m}(k_mr)J_\lambda(k_Br)e^{i\,(\lambda-m)\,\varphi} \ud\mr \nonumber \\
& \times \int_{D_n} J_{n}(k_nr')J_\lambda(k_Br')e^{i\,(\gamma+n)\,\varphi'} \ud\mr'.
\end{align}

Due to the circular symmetry, the angular integrations over the domains $D_m$ and $D_n$ lead to non-zero values only for $\lambda=m$ and $\gamma=n$. In these cases the radial integrals have well known analytical values, leaving only a final integration over $k_x$.

\subsection{Light emission in finite sized photonic crystal waveguide}
\label{Sec:PhotonGunTM}
\begin{figure}[htb]
\centering
\includegraphics{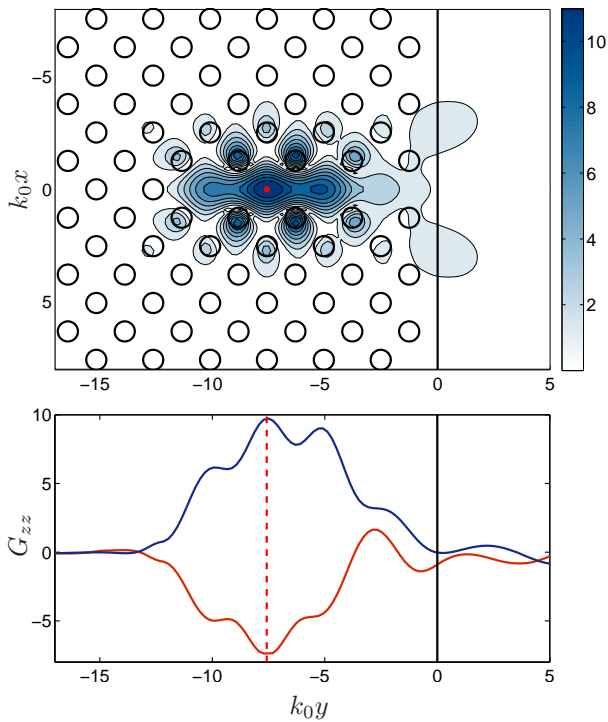}
\caption{\label{Fig:MekisPhGun_WithInterface_montage}Top: Absolute value $|G_{zz}(\mr,\mr')|$ of the TM Green's tensor for a finite sized photonic crystal waveguide consisting of 80 rods of refractive index $n_d=3.4$ in a background with an interface between a low-index dielectric ($n_B=1.5$) and air ($n_A=1$). The results are calculated as function of $\mr$ with $k_0\mr'=(0,-7.58)$ (indicated by the red dot and vertical dashed line). Bottom: Real (red) and imaginary (blue) parts of $G_{zz}(y,\mr')$ along the line $x=0$.}
\end{figure}
\begin{figure}[htb]
\centering
\includegraphics{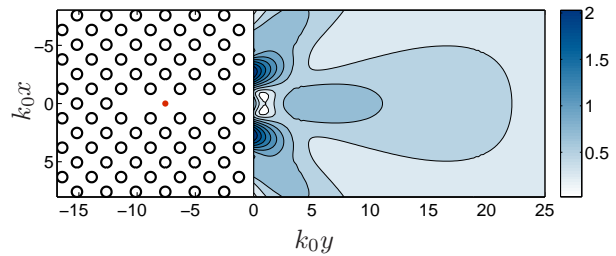}
\caption{\label{Fig:MekisPhGun_emissionAbsContourWithInterface}Contour plot of emission pattern, $|G_{zz}(\mr,\mr')|$, of the system in Fig. \ref{Fig:MekisPhGun_WithInterface_montage}, but for positions outside the photonic crystal.}
\end{figure}

In Fig. \ref{Fig:MekisPhGun_WithInterface_montage} we show a contour plot of the absolute value of the Green's tensor $|G_{zz}(\mr,\mr')|$ along with real and imaginary parts at positions along the $x$-axis. Results are shown for $k_0\mr'=(0,-7.58)$, in the center of the waveguide at the location of one of the missing rods. In an infinite waveguide, this would be the location of the field antinode of the waveguide mode. The periodic Bloch-mode character of the waveguide mode is evident also in the case of this finite waveguide and the structure acts as a resonator, greatly increasing the absolute value of the Green's tensor for positions $\mr$ inside the waveguide as compared to the bulk medium. For $\mr\rightarrow\mr'$ the real part of $G_{zz}(\mr,\mr')$ diverges. This is the case also in Fig. \ref{Fig:MekisPhGun_WithInterface_montage}, but the divergence is too weak to show up at the chosen discretization. Although the finite waveguide acts as a resonator, light can propagate out of the end facet. Fig. \ref{Fig:MekisPhGun_emissionAbsContourWithInterface} shows $|G_{zz}(\mr,\mr')|$ at positions outside the structure. As noted in section \ref{Sec:BGelectricField}, the Green's tensor is related to the electric field at point $\mr$ due to a line source at point $\mr'$. Therefore we may interpret the figure as the emission pattern from the source inside the waveguide. Due to the resonator effect of the waveguide structure, the emission pattern does not show up on the scale of the contour plot in Fig. \ref{Fig:MekisPhGun_WithInterface_montage}.

%

\section{Conclusion}
\label{Sec:Conclusion}
We have described a procedure for solving the Lippmann-Schwinger equation for electromagnetic scattering in which the field along with the electric field Green's tensor is expanded in a basis of cylinder functions (so-called normal modes) inside each scatterer. Projection of the electric field and the Green's tensor onto the normal modes is facilitated by the use of a number of addition theorems to simplify the integral expressions and we have presented the method in general along with a thorough discussion of the evaluation of the scattering matrix elements, which may be helpful for practical implementation.

The basis of normal modes ensures that all basis functions have the correct wave number. This, combined with the need for solving the system inside the scattering elements only, results in a relatively small linear equation system, as compared with other methods. Consequently, the method is fast and capable of handling large material systems such as photonic crystals. Furthermore, the use of a local cylinder function basis avoids the introduction of fictitious charges which may lead to instabilities for large refractive index contrasts in the case of TE polarization \cite{PetersonRayMittra} and the integration scheme is free of staircasing errors along the boundaries. Due to the formulation in terms of the Green's tensor of the background medium, there is no need for a calculation domain and the radiation condition is automatically satisfied, as are the boundary conditions (limited only by the numerical precision chosen). The accuracy of the method is thus limited only by the number of basis functions and the tolerance on the numerical integrals employed for the evaluation of the scattering matrix elements. We have introduced a measure of accuracy based on self-consistency and we have shown it to be of practical as well as principal importance. Once the matrix equation has been set up, it holds all information necessary to carry out scattering calculations on the geometry at the chosen frequency. It can thus be stored and used for different choices of incoming fields as well as for the calculation of the Green's tensor between different points $\mr$ and $\mr'$.

We have illustrated the method by two example problems and we have shown an application of the method where we have calculated the $zz$ component of the Green's tensor of a finite sized photonic crystal waveguide. Similar calculations will find application in the development of nanophotonic devices such as in the design of junctions or cavities in photonic crystals or investigation of the emission pattern from single photon sources. Using a similar procedure as the one described the method may be extended to three dimensional scattering geometries and although we have focused on applications in micro- and nano photonic structures, we believe the method may be of use in other areas of electromagnetic scattering calculations as well.

\section*{Acknowledgement}
This work was supported by the Villum Kann Rasmussen Centre of Excellence 'NATEC'.

\appendix

\section{Addition theorems for multipole expansions}
\label{Sec:AdditionTheorems}
The expansion of the Lippmann-Schwinger equation used in this work, and especially the calculation of matrix elements, rely heavily on the use of cylinder functions. A number of addition theorems exist for cylinder functions which may simplify the calculations considerably. Of special interest in the present work is the Jacobi-Anger identity and the Graf addition theorem \cite{Erdelyi_II}. Below we summarize the results in forms suitable for the present application.

\subsection{Jacobi-Anger identity}
For a plane wave, travelling at an angle $\theta$ with respect to the $x$ axis, the expansion in terms of cylinder functions is given by the Jacobi-Anger identity
\begin{equation}
e^{i\,k_0r\,\cos(\varphi-\theta)} = \sum_{n=-\infty}^{\infty} i^ne^{-i\,n\,\theta}J_n(k_0r)e^{i\,n\,\varphi},
\label{Eq:JacobiAnger}
\end{equation}
in which $(r,\varphi)$ are cylindrical coordinates and $J_n(k_0r)$ is the Bessel function of order n.

\subsection{Graf's addition theorem}
The Graf addition theorem may be used to express the cylinder functions in one local coordinate system in terms of cylinder functions in a different local coordinate system.
\begin{figure}[htb]
\centering
\includegraphics{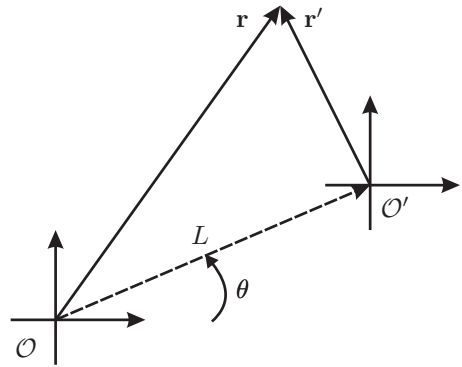}
\caption{\label{Fig:GrafAdditionTheorem}Sketch of relative coordinates as used in the expression for Graf's addition theorem.}
\end{figure}

We consider the cylindrical coordinates $\mr=(r,\varphi)$ and $\mr'=(r',\varphi')$, centered at two different positions $\mathcal{O}$ and $\mathcal{O}'$, respectively, where $(L,\theta)$ denotes the coordinates of $\mathcal{O}'$ with respect to $\mathcal{O}$ as shown in Fig. \ref{Fig:GrafAdditionTheorem}. Using this notation we express the Graf addition theorem as
\begin{equation}
Z_n(k r)e^{in(\varphi-\theta)} = \sum_{\mu=-\infty}^\infty Z_{n+\mu}(k L)Z_{\mu}(k r')(-1)^\mu e^{i\mu(\theta-\varphi')},
\label{Eq:GrafAdditionTheorem}
\end{equation}
where $Z_n$ is a solution to Bessel's differential equation for integer $n$. If $Z_n=J_n$, the expansion is valid for all values of $r'$, otherwise it is valid only for $r'<L$.

\section{Calculation of Matrix elements}
\label{Sec:MatrixElementCalculations}
In this appendix we evaluate the integral
\[
I^{\alpha\beta}_{\mu}=\int_{P-\delta V}G^B_{\alpha\beta}(\mr,\mathbf{R})J_\mu(k_RR)(-1)^{\mu}e ^{-i\mu\theta} \ud \mathbf{R},
\]
which enters the expression for $\mathcal{A}^{\alpha\beta}$ in Section \ref{Sec:SelfTerms}.

For TM polarization, $(\alpha,\beta)=(z,z)$, the angular integration is nonzero only for $\mu=0$ and the resulting integral is well behaved, allowing for easy evaluation. For TE polarization, the integrand has a pole at the origin, so we rewrite this integral in a form more suitable for numerical quadrature. Although the procedure is the same, the resulting integrals differ slightly depending on which of the elements of the Green's tensor we consider (see Ref. \cite{Martin_PRE58_3909_1998} for explicit expressions for the elements). For $I_\mu^{xx}$ we get
\begin{align}
I_\mu^{xx} &=\frac{i}{4}\int_0^\infty\int_0^{2\pi}\left\{\sin^2\theta H_0(k_BR) + \frac{\cos(2\theta)}{k_BR}H_1(k_B(R) \right. \nonumber \\
&\quad + \left. \frac{2i}{\pi}\frac{\cos(2\theta)}{k_B^2R^2} \right\} J_\mu(k_RR)(-1)^\mu e^{-i\mu\theta}R\ud \theta\ud R \nonumber \\
&\quad - \frac{i}{4} \lim_{\delta\rightarrow 0} \int_\delta^\infty\int_0^{2\pi} \frac{2i}{\pi}\frac{\cos(2\theta)}{k_B^2R^2} J_\mu(k_RR)(-1)^\mu e^{-i\mu\theta}R\ud \theta\ud R \nonumber \\
&=\frac{i\pi}{4}\int_0^\infty \left\{ (-\frac{1}{2}\delta_{\mu,-2} + \delta_{\mu,0} -\frac{1}{2}\delta_{\mu,2} ) H_0(k_BR) \right.\nonumber \\
&\quad + \left. ( \delta_{\mu,-2} + \delta_{\mu,2}) ( \frac{H_1(k_BR)}{k_BR} + \frac{2i}{\pi k_B^2R^2} )\right\} J_\mu(k_RR)\,R\ud R \nonumber \\
&\quad \frac{1}{2}\lim_{\delta\rightarrow 0}\int_\delta^\infty (\delta_{\mu,-2}+\delta_{\mu,2})\frac{J_\mu(k_RR)}{k_B^2R^2}R\ud R \nonumber
\end{align}

The first integral is now well behaved and may be directly evaluated, whereas for the second integral we may use the identity
\[
\lim_{\delta\rightarrow 0}\int_\delta^\infty\frac{J_2(K_RR)}{k_B^2R^2} R\ud R = \frac{1}{2k_B}.
\]

In a similar way we rewrite the expressions for $I_\mu^{yy}$ and $I_\mu^{xy}$ as follows:
\begin{align}
I_\mu^{yy} &=\frac{i\pi}{4}\int_0^\infty \left\{ (\frac{1}{2}\delta_{\mu,-2} + \delta_{\mu,0} \frac{1}{2}\delta_{\mu,2} ) H_0(k_BR) \right.\nonumber \\
&\quad - \left. ( \delta_{\mu,-2} + \delta_{\mu,2}) ( \frac{H_1(k_BR)}{k_BR} + \frac{2i}{\pi k_B^2R^2} )\right\} J_\mu(k_RR)\,R\ud R \nonumber \\
&\quad -\frac{1}{4k_B^2}(\delta_{\mu,-2}+\delta_{\mu,2})\nonumber .
\end{align}

\begin{align}
I_\mu^{xy} &=-\frac{\pi}{4}\int_0^\infty ( \delta_{\mu,-2} - \delta_{\mu,2}) ( \frac{1}{2}\frac{H_2(k_BR)}{k_BR} + \frac{2i}{\pi k_B^2R^2} ) J_\mu(k_RR)\,R\ud R \nonumber \\
&\quad +\frac{i}{4k_B^2}(\delta_{\mu,-2}-\delta_{\mu,2})\nonumber .
\end{align}

\section{Derivatives for general cylinder functions}
\label{Sec:GeneralDerivatives}
Below we summarize in polar coordinates, the double derivatives of the general cylinder function $Z_\lambda(kr)\exp\{\pm i\lambda\phi\}$, in which $(r,\phi)$ are cylindrical coordinates and $Z_\lambda$ is a solution to Bessels differential equation for integer $\lambda$.
\begin{align}
\frac{\partial^2}{\partial x^2}&\left\{Z_\lambda(k r)e^{i \lambda \varphi} \right\} = \frac{k ^2}{4}\left\{ Z_{\lambda+2}(kr)e^{i(\lambda+2)\varphi} \right. \nonumber \\
& \left. + Z_{\lambda-2}(kr)e^{i(\lambda-2)\varphi} - 2 Z_\lambda(kr)e^{i\lambda\varphi}\right\},
\end{align}

\begin{align}
\frac{\partial^2}{\partial y^2}&\left\{Z_\lambda(k r)e^{i \lambda \varphi} \right\} = -\frac{k ^2}{4}\left\{ Z_{\lambda+2}(kr)e^{i(\lambda+2)\varphi} \right. \nonumber \\
& \left. + Z_{\lambda-2}(kr)e^{i(\lambda-2)\varphi} + 2 Z_\lambda(kr)e^{i\lambda\varphi}\right\},
\end{align}

\begin{align}
\frac{\partial^2}{\partial x\partial y}\left\{Z_\lambda(k r)e^{i \lambda \varphi} \right\} &= -i\frac{k ^2}{4}\left\{Z_{\lambda+2}(kr)e^{i(\lambda+2)\varphi} \right. \nonumber \\
& \left. - Z_{\lambda-2}(kr)e^{i(\lambda-2)\varphi} \right\}.
\end{align}

%
%
%
%
%


\end{document}